\documentclass{PoS}
\pdfoutput=1

\usepackage{amsmath}

\newcommand{\mustbe}{\stackrel{!}{=}}
\newcommand{\of}[1]{\left(#1\right)}
\newcommand{\at}[1]{\left.#1\right\vert}
\newcommand{\eq}[1]{\begin{align}#1\end{align}}
\newcommand{\vev}[1]{\left\langle#1\right\rangle}
\newcommand{\der}[3]{\frac{d^{#1}{#3}}{d{#2}^{#1}}}
\newcommand{\pder}[3]{\frac{\partial^{#1}{#3}}{\partial{#2}^{#1}}}
\newcommand{\subeq}[1]{\begin{subequations}\eq{#1}\end{subequations}}

\usepackage{textpos}
\newcommand{\preprint}[1]{
	\begin{textblock}{1}(0,-3.1)
		\texttt{\hfill\footnotesize #1}
	\end{textblock}
}

\title{Thermodynamics of strongly-coupled lattice QCD\\in the chiral limit
\preprint{CERN-PH-TH-2017-012}
}

\ShortTitle{Thermodynamics of strongly-coupled lattice QCD in the chiral limit}

\author{Philippe de Forcrand\\
        Institut f\"{u}r Theoretische Physik, ETH Z\"{u}rich, CH-8093 Z\"{u}rich, Switzerland\\
        CERN, TH Division, CH-1211 Geneva 23, Switzerland\\
        E-mail: \email{forcrand@phys.ethz.ch}}

\author{Paul Romatschke\\
        Department of Physics, 390 UCB, University of Colorado at Boulder, Boulder, CO, USA\\
        E-mail: \email{paul.romatschke@colorado.edu}}

\author{Wolfgang Unger\\
        Fakult\"at f\"ur Physik, Universit\"at Bielefeld, 
        Universit\"atstasse 25, D33619 Bielefeld, Germany\\
        E-mail: \email{wunger@physik.uni-bielefeld.de}}

\author{\speaker{H\'{e}lvio Vairinhos}\\
        Institut f\"{u}r Theoretische Physik, ETH Z\"{u}rich, CH-8093 Z\"{u}rich, Switzerland\\
        E-mail: \email{helviov@phys.ethz.ch}}

\abstract{In the strong coupling limit, $n$-point functions in lattice QCD with staggered fermions can be rewritten exactly as sums over constrained configurations of monomers, dimers, and baryon loops covering the spacetime lattice. Worm algorithms provide efficient global sampling methods over such ensembles, and are particularly efficient in the chiral limit. We study the thermodynamics of strongly-coupled $U(3)$ and $SU(3)$ lattice QCD with one massless staggered fermion using such methods, and compare the results with the relativistic pion gas down to low temperatures $O(15~ {\rm MeV})$.}

\FullConference{34th annual International Symposium on Lattice Field Theory\\
		24-30 July 2016\\
		University of Southampton, UK}

\begin{document}

%------------------------------------------------------------------------------------------------------------
\section{Introduction}

Consider $U(N)$ or $SU(N)$ lattice QCD with a single staggered fermion flavour, at finite temperature. 
At low temperatures, the chiral U(1) symmetry of the massless staggered fermion is spontaneously broken, to which a single massless Goldstone boson is associated: the pion. 

The pion is interacting, but at sufficiently low temperatures the strength of the effective interactions vanishes, i.e. $\frac{T}{F_\pi} \to 0$, and the pion is effectively free. In such a regime, the physics is that of an ideal pion gas, whose energy density, $\epsilon$, satisfies the Stefan-Boltzmann law (SB) for a single bosonic degree of freedom:
\eq{
	\varepsilon(T) = \varepsilon(0) + \frac{\pi^2}{30} T^4
}

Here we summarize our numerical study of the thermal properties of $U(3)$ and $SU(3)$ lattice QCD with a single staggered fermion, in the chiral limit, where we test the hypothesis of a (near) ideal pion gas below the critical temperature of the chirally-restoring phase transition. 

We choose to perform simulations in the strong coupling limit, $\beta = 0$, for there we have access to Monte Carlo algorithms of the worm type, which are very efficient, even in the chiral limit and at low temperatures. This allows us to determine the equation of state of lattice QCD with high precision, at unprecedentedly low temperatures.

%------------------------------------------------------------------------------------------------------------
\section{Thermodynamics of a free massless boson on the lattice\label{sec:karsch}}

First, it is instructive to understand the behavior of an ideal gas of massless bosons on a lattice. Lattice corrections to the ideal gas regime of a free massless boson, on a $N_s^3 \times N_t$ lattice with anisotropy $\xi = {a}/{a_t}$, have been studied in \cite{Engels:1981ab}. The energy density $\varepsilon$ of such a gas is given by:
\subeq{
a^4 \varepsilon(T) &= 
-\frac{\xi^3}{N_s^3 N_t} \sum_{\vec\jmath\neq \vec 0} 
\frac{\sin^2\of{{\pi j_0}/{N_t}}}
{b^2 + \xi^2 \sin^2\of{{\pi j_0}/{N_t}}},
\qquad
b^2 = \sum_{i=1}^3\sin^2\of{{\pi j_i}/{N_s}}
\\
a^4 \varepsilon(0) &= - \frac{\xi^3}{N_s^3} \sum_{\vec\jmath\neq \vec 0} 
\of{b^2 + \xi^2 + b\sqrt{b^2 + \xi^2}}^{-1}
}
where the lattice temperature is given by $aT = {\xi}/{N_t}$. Similar expressions can be obtained for the pressure $p$. In particular, they imply that the trace anomaly vanishes on any finite lattice \cite{Engels:1981ab}: 
\eq{
\Delta\varepsilon - 3\Delta p = 0
\label{eq:trace-anomaly}
}
where $\Delta\varepsilon(T) = \varepsilon(T) - \varepsilon(0)$, and $\Delta p(T) = p(T) - p(0)$.

\begin{figure}[h!]
\centering
\includegraphics[scale=.32]{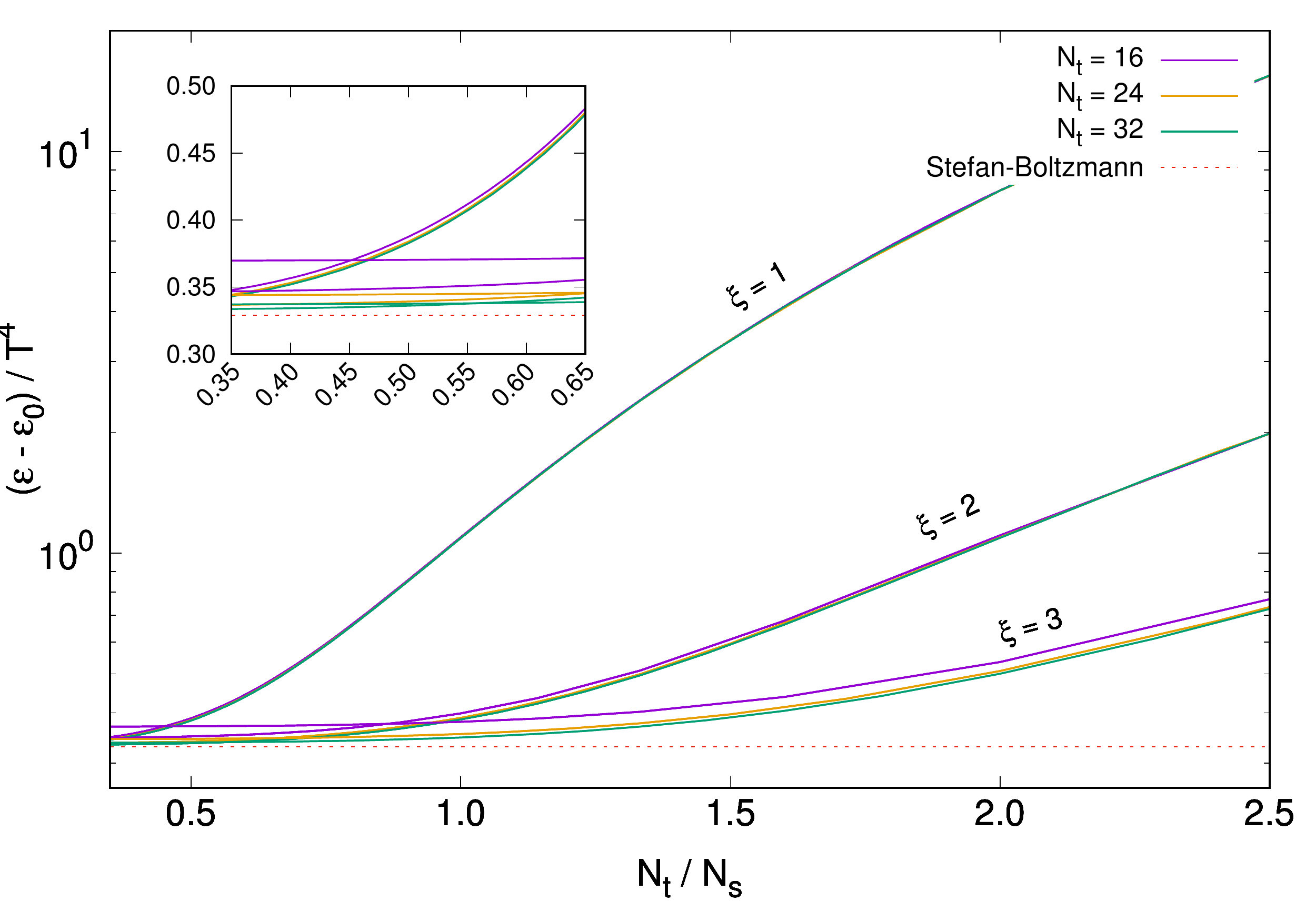}
\vskip -2mm
\caption{Energy density of a free massless boson on the lattice: the finite-size effects induce very large deviations from the Stefan-Boltzmann limit (dotted line), but are significantly suppressed (less than 10\%) when $N_s \gtrsim 3 N_t$ for $\xi = 1$, or $N_s \gtrsim 2 N_t$ for $\xi \geq 2$.}
\label{fig:ideal-gas}
\end{figure}

In this system, discretization effects are quite significant (see Fig.~\ref{fig:ideal-gas}): from \cite{Engels:1981ab}, we learn that lattice corrections are small for $N_s \geq 2 N_t$ and $\xi\geq 2$. Actually, the ideal gas behavior is only exact in the continuous time limit, $\xi\to\infty$.

We keep this in mind when simulating lattice QCD in the regime where the $U(1)$ chiral symmetry is spontaneously broken, and the pion is massless. Even though pions are not free ($F_\pi \neq 0$), their interactions should be negligible in the regime $T\ll F_\pi$ (or in the large $N$ limit), and the picture of an ideal pion gas should become a good approximation.

%------------------------------------------------------------------------------------------------------------
\section{Dimer representation of lattice QCD in the strong coupling limit}

The partition function of $SU(3)$ lattice QCD with $N_f=1$ staggered fermions, at $\beta = 0$, is: 
\eq{
Z &=\int {\cal D}U {\cal D}\psi {\cal D}\bar\psi\; e^{2a_t m_q \sum_x \bar\psi_x\psi_x
	+\sum_{x,\mu} \gamma^{\delta_{\mu 0}} \eta_{x\mu} \of{e^{a_t\mu_q}\bar\psi_x U_{x\mu} \psi_{x+\hat\mu} - e^{-a_t\mu_q}\bar\psi_{x+\hat\mu} U_{x\mu}^\dag \psi_x}}
	\label{eq:partition-SU3}
}
where $\mu_q$ is the quark chemical potential, $\gamma$ is the bare anisotropy, $m_q$ is the bare quark mass, $a_t$ ($a$) is the temporal (spatial) lattice spacing, and $\eta_{x\mu} = \pm 1$ are the staggered phases.

Analytic integration of the link variables, followed by the integration of the Grassmann variables, yields the partition sum of a system of monomers, dimers, and baryon loops \cite{mdp}:
\eq{
Z &= \sum_{\{n,k,\ell\}} \frac{\sigma(\ell)}{3!^{\vert \ell \vert}} \of{\prod_x \frac{3!}{n_x!}}
\of{\prod_{x,\mu} \frac{(3 - k_{x\mu})!}{3! k_{x\mu}!}}
(2a_t m_q)^{N_M} \gamma^{2N_{Dt} + 3N_{\ell t}}  
e^{3N_t a_t \mu_q w_\ell}
\label{eq:partition-mdp-SU3}
}
where $n_x,k_{x\mu} \in \{0,1,2,3\}$ are occupation numbers of monomers and dimers, $\ell_{x\mu} \in \{0,\pm 1\}$ are occupation numbers of oriented baryonic dimers, and $N_M, N_{Dt}, N_{\ell t}$ denote, respectively, the total number of monomers, timelike dimers, and timelike baryonic links:
\eq{
N_{M} = {\sum}_x n_x, 
\qquad 
N_{Dt} = {\sum}_x k_{x0}, 
\qquad 
N_{\ell t} = {\sum}_x \vert \ell_{x0} \vert
}
$w_\ell$ counts the number of times baryon loops wrap around the thermal direction, and $\sigma(\ell)=\pm 1$ is a sign which depends on the shape of the baryon loops (and introduces a sign problem).

Due to the Grassmann integration, the configurations which contribute to the partition function are constrained, on each site, to have either exactly 3 monomers and/or dimers, or be traversed by a non-self-intersecting oriented baryon loop:
\eq{
	n_x + {\sum}_{\pm\mu} k_{x\mu} \mustbe 3,
	\qquad
	{\sum}_{\pm\mu} \ell_{x\mu} \mustbe 0,
	\qquad
	\forall x
	\label{eq:grassmann}
}

Such constrained configurations can be efficiently sampled using variants of the worm algorithm: a ``mesonic worm'', which updates the monomer-dimer sector \cite{worm}, and a ``baryonic worm'', which updates the baryonic loops and the 3-dimer sector \cite{deForcrand:2009dh,Unger:2011in}.

The partition function for $U(3)$ QCD is obtained from \eqref{eq:partition-mdp-SU3} by removing the baryons, i.e. $\ell_{x\mu} = 0, \forall x,\mu$. In this case, only the mesonic worm is needed to simulate it.

%------------------------------------------------------------------------------------------------------------
\section{Thermodynamics of lattice QCD in the strong coupling and chiral limits}

The energy density and pressure in $SU(3)$ lattice QCD, in the strong coupling limit ($\beta=0$) and chiral limit ($m_q=0$), are related to the density of hadrons hopping in the time direction (i.e. timelike dimers and timelike baryon links):
\subeq{
	a^3 a_t\, \Delta\varepsilon
	= 
	\mu_B \rho_B
	-\frac{a^3 a_t}{V} \at{\pder{}{T^{-1}}{\log Z}}_{V,\mu_B}
	&= \frac{\xi}{\gamma} \der{}{\xi}{\gamma} \vev{2 n_{Dt} + 3n_{\ell t}}
	\label{eq:energy}
	\\
	a^3 a_t\, \Delta p 
	= a^3 a_t T \at{\pder{}{V}{\log Z}}_{T,\mu_B}
	&= \frac{\xi}{3\gamma} \der{}{\xi}{\gamma} 
	\vev{2 n_{Dt} + 3n_{\ell t}}
	\label{eq:pressure}
}
where $\mu_B$ is the baryon chemical potential, $\rho_B = \vev{w_\ell}/N_s^3$ is the baryon density, and $\xi(\gamma) = \frac{a}{a_t}$ is the renormalized anisotropy, which parameterizes the physical anisotropy of the lattice. In order to obtain the corresponding expressions in $U(3)$ QCD, it suffices to take $\rho_B=0$ and $n_{\ell t}=0$.

Eqs. \eqref{eq:energy} and \eqref{eq:pressure} imply that the trace anomaly vanishes for any lattice spacing, cf. \eqref{eq:trace-anomaly}.

%------------------------------------------------------------------------------------------------------------
\section{Anisotropy calibration}

An accurate determination of the energy density \eqref{eq:energy}, or pressure \eqref{eq:pressure}, requires a precise knowledge of the renormalized anisotropy $\xi$ as a function of the bare anisotropy $\gamma$, and also of its running, $d\xi/d\gamma$. For this purpose, we use the fluctuations of certain conserved charges, labelled by spacetime directions, as probes for the calibration of the lattice anisotropy.

In the chiral limit, Grassmann constraints \eqref{eq:grassmann} imply the existence of conserved currents \cite{Chandrasekharan:2006tz}:
\eq{
j_{x\mu} = \pi_x \of{k_{x\mu} - \frac{3}{2}\vert \ell_{x\mu} \vert - \frac{3}{8}}
\quad \Rightarrow \quad
\sum_{\pm\mu} j_{x\mu}= 0, \; \forall x
}
where $\pi_x = (-1)^{\sum_\mu x_\mu} = \pm 1$ is the parity (bipartite color) of the site $x$.
We can also define conserved charges by integrating the currents along codim-1 hyperslices ${\cal S}_\mu$ perpendicular to the direction $\hat\mu$:
\eq{
Q_\mu = \sum_{x \in {\cal S}_\mu} j_{x\mu}
}
Due to parity symmetry, $\vev{Q_\mu}=0, \;\forall\mu$. We consider lattices with the same size $N_s$ in all spatial directions, and thus compare variations of the timelike charge, $Q_t^2 = Q_0^2$, and of the average of the spacelike charges, $Q_s^2 = \frac{1}{3}\sum_{i=1}^3 Q_i^2$.

\begin{figure}[t]
\centering
\includegraphics[scale=.28]{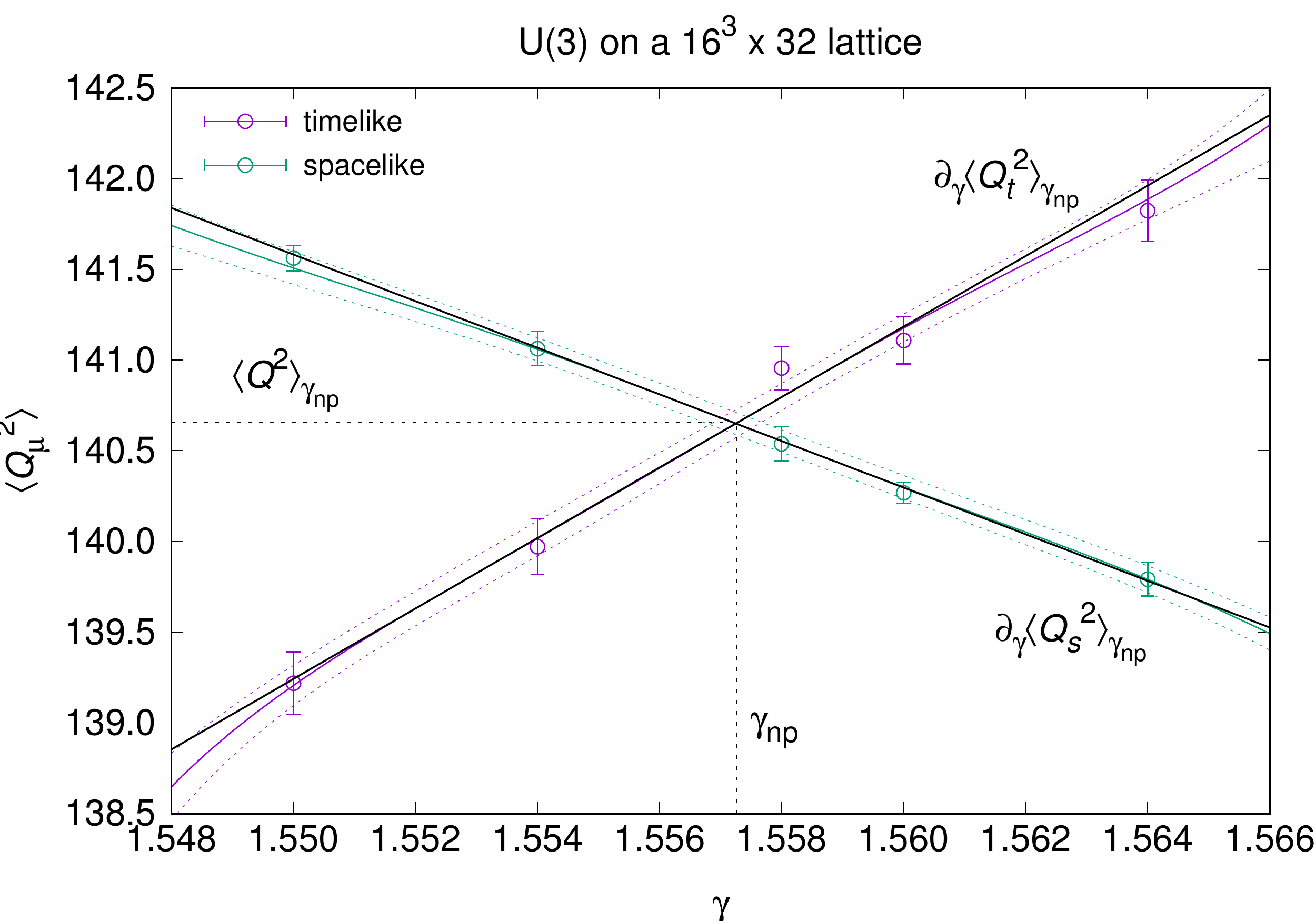}
\vskip -2mm
\caption{Fluctuations of the conserved charges in the timelike (purple) and spacelike (green) directions, as a function of the bare anisotropy $\gamma$, in $U(3)$ QCD. The intersection point corresponds to the critical value of $\gamma$ for which the physical box is hypercubic, while the lattice has an anisotropy $\xi(\gamma_{\rm np}) = \frac{N_t}{N_s} = 2$.}
\label{fig:intersect}
\end{figure}

Our non-perturbative renormalization criterion requires the fluctuations of the conserved charges to be isotropic when the physical volume is hypercubic (in the thermodynamic limit):
\eq{
\vev{Q_t^2}({\gamma_{\rm np}}) &= \vev{Q_s^2}({\gamma_{\rm np}})
\quad\Rightarrow\quad
\frac{aN_s}{a_t N_t} = \xi(\gamma_{\rm np}) \frac{N_s}{N_t} = 1
}
where $\gamma_{\rm np}$ is the nonperturbative, finely tuned value of the bare anisotropy for which fluctuations of the conserved charges coincide. Using scaling arguments, it is also easy to relate the running of the renormalized anisotropy to expectation values associated with these conserved charges:
\eq{
\xi \der{}{\xi}{\gamma} = \frac{\vev{Q^2}_{\gamma_{\rm np}}}{\at{\of{\der{}{\gamma}{}{\vev{Q_t^2}} - \der{}{\gamma}{}{\vev{Q_s^2}}}}_{\gamma_{\rm np}}}
}
where $\vev{Q^2}_{\gamma_{\rm np}}$ is the variance at $\gamma = \gamma_{\rm np}$, and the denominator depends only on the difference of their slopes at $\gamma = \gamma_{\rm np}$. We determine all these quantities by finding the intersection point of the curves of $Q_t^2$ and $Q_s^2$, which are constructed using multi-histogram reweighting (Fig.~\ref{fig:intersect}).

\begin{figure}[t]
\centering
\includegraphics[scale=.28]{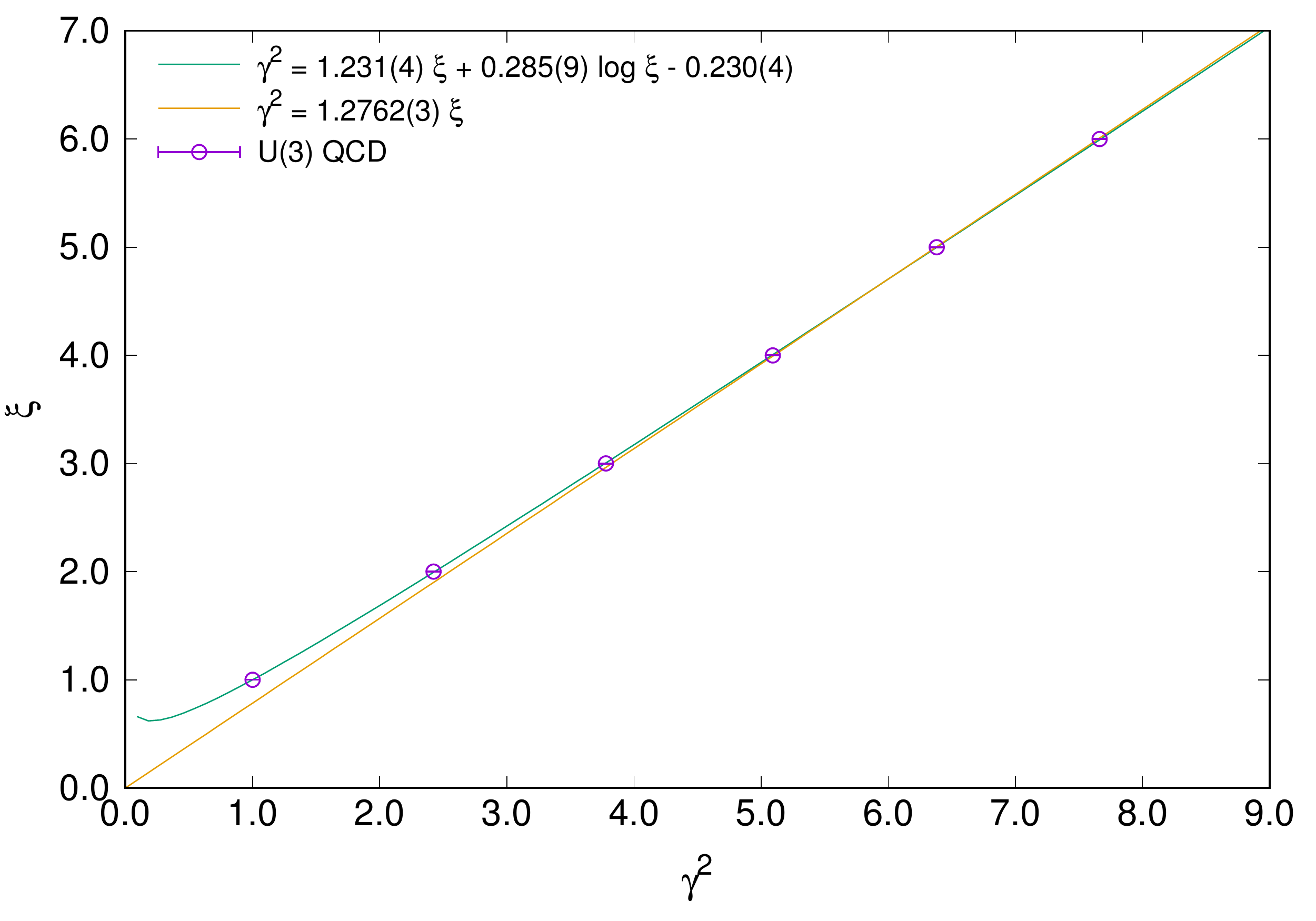}
\includegraphics[scale=.28]{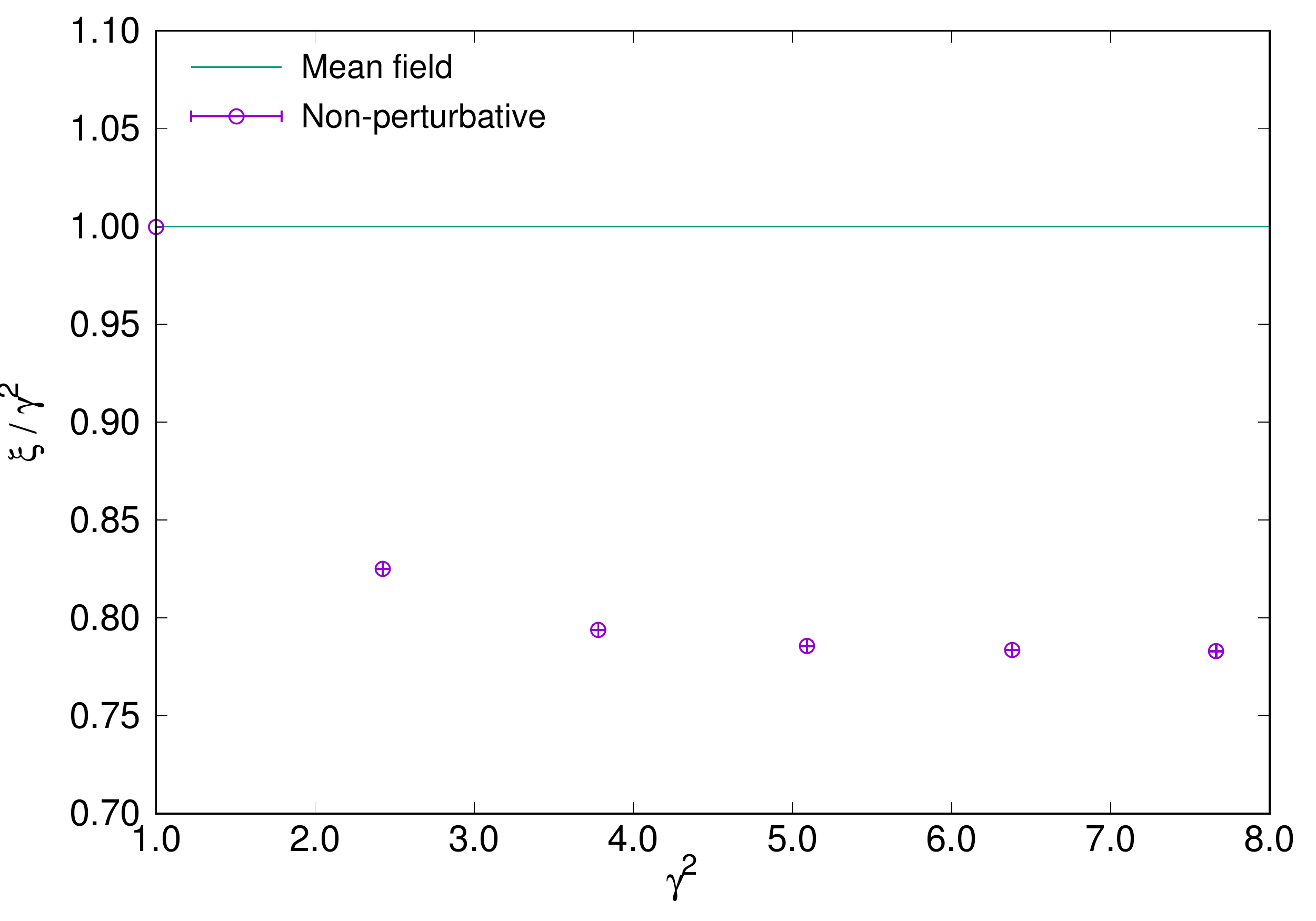}
\vskip -2mm
\caption{Nonperturbative relation between the bare and renormalized anisotropy in $U(3)$ QCD (left), and its deviation from the mean field prediction (right).}
\label{fig:aniso}
\end{figure}

We determine $\gamma_{\rm np}$ for several aspect ratios, $ \frac{N_t}{N_s} = \xi \in \{2,3,4,5,6\}$, and for different spatial sizes, in $U(3)$ and $SU(3)$ QCD. In the thermodynamic limit, the functional dependence $\xi(\gamma)$ appears to be quadratic for large $\gamma$ (Fig.~\ref{fig:aniso}, left). Mean field arguments also predict a quadratic dependence in the large $\gamma$ limit: $\xi(\gamma) = \gamma^2$ \cite{aniso-mean-field}, but the non-perturbative prefactor differs from the mean-field one by $\approx 25\%$ (Fig.~\ref{fig:aniso}, right).

%------------------------------------------------------------------------------------------------------------
\section{Energy density vs. temperature}

Given the relation between the bare and renormalized anisotropies, and the corresponding running (Fig.~\ref{fig:aniso}), the remaining ingredient for an accurate determination of the energy density \eqref{eq:energy} in $U(3)$ QCD is a precise measurement of the density of timelike dimers, $n_{Dt}$. 

In order to determine the dependence of the energy density $\varepsilon$ on the temperature $T$, we first need to accurately subtract from it the $T=0$ contribution, $\varepsilon_0$.
We compute $\varepsilon_0$ by taking the thermodynamic limit of the density of timelike dimers evaluated on a hypercubic lattice: 

\begin{figure}[t]
\centering
\includegraphics[scale=.295]{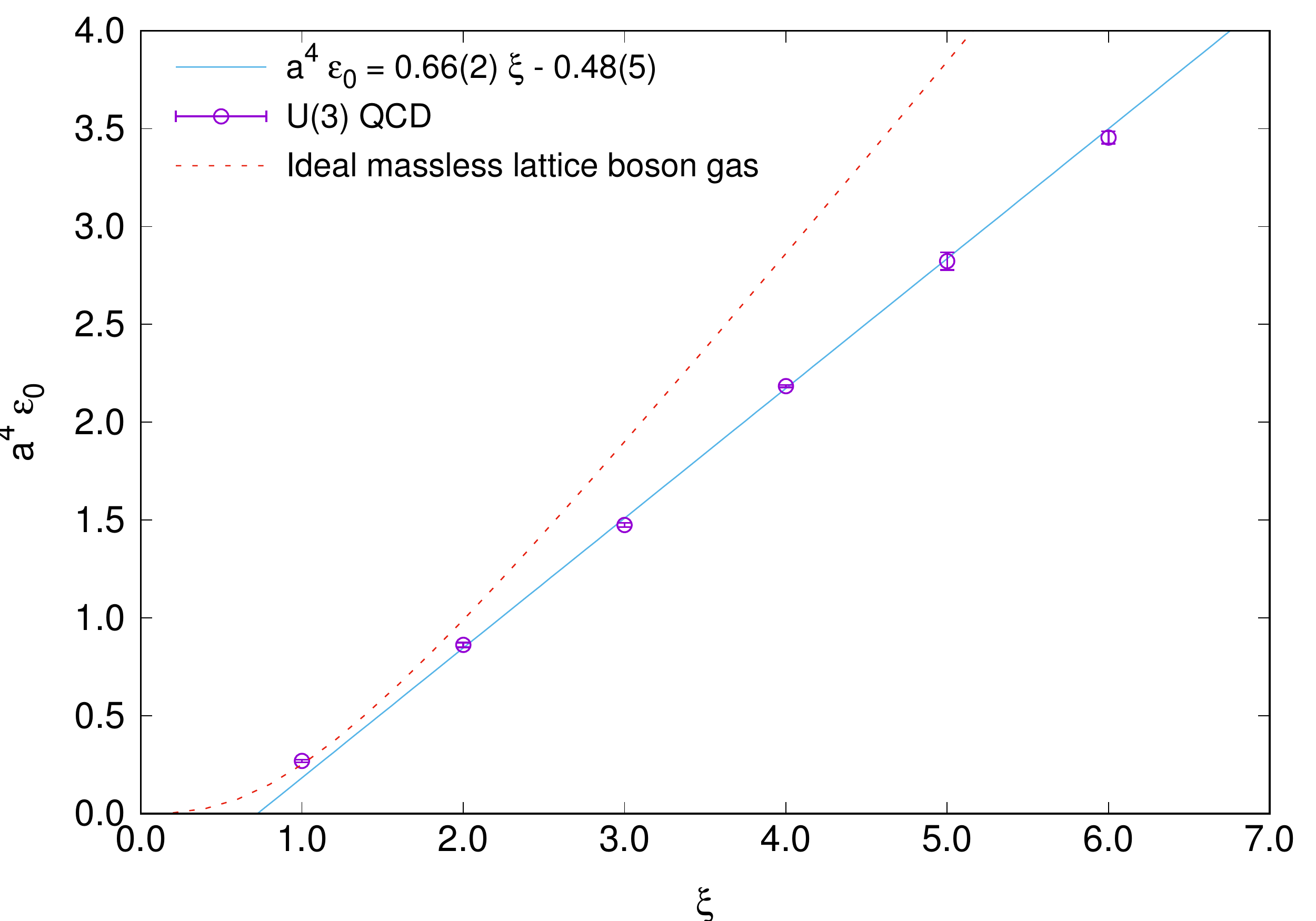}
\includegraphics[scale=.295]{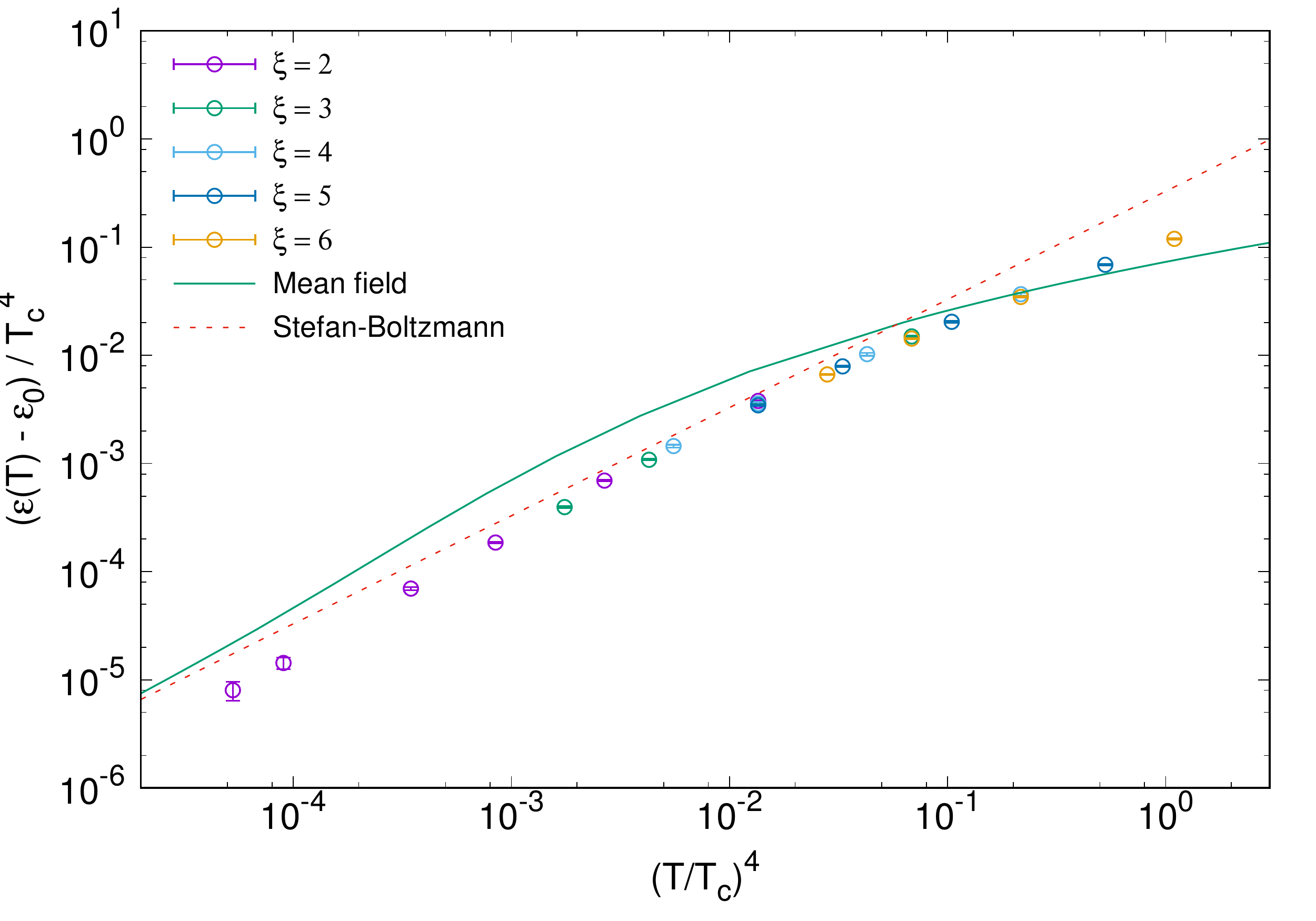}
\vskip -2mm
\caption{Scaling of the energy density with the physical anisotropy, at $T=0$ (left), and the subtracted energy density $(\varepsilon(T) - \varepsilon_0)$ as a function of the temperature (right), both for $U(3)$ QCD. There are significant deviations to the equation of state of an ideal pion gas, for low temperatures and for temperatures near $T_c$. Note that the lowest temperature reached is $O(15~{\rm MeV})$.}
\vskip -9mm
\label{fig:energy}
\end{figure}

\eq{
a^4\varepsilon_0 (\xi) = \lim_{N_s\to\infty} \at{\frac{\xi^2}{\gamma} \der{}{\xi}{\gamma} \vev{2 n_{Dt}}}_{N_t=\xi N_s}
}
We observe that $\varepsilon_0$ scales approximately linearly with $\xi$, for large $\xi$ (Fig.~\ref{fig:energy}, left), similarly to an ideal gas of massless scalar bosons on the lattice \cite{Engels:1981ab}, but with a different non-universal prefactor.

At finite temperature, we compute the energy density $a^4 \Delta\varepsilon(N_s,N_t,\xi)$, with the $\varepsilon_0$ contribution subtracted, on $N_s^3 \times N_t$ lattices, for fixed $aT = \frac{\xi}{N_t}$, and for several spatial sizes. We then take the thermodynamic limit of $a^4 \Delta\varepsilon$, assuming $O(N_s^{-3})$ corrections.\footnote{~Inspired by the lessons of the ideal gas of massless scalar bosons on the lattice (Section \ref{sec:karsch}), we take the thermodynamic limit, $N_s\to\infty$, by only using lattices for which $N_s \geq 2 N_t$, in order to minimize the finite-size corrections.} We express both energy density and temperature in units of the critical temperature of the chiral phase transition, which for $U(3)$ is $aT_c = 1.466$ and for $SU(3)$ is $aT_c = 1.089$.\footnote{~In the literature, the values of the critical temperature, namely $aT_c = 1.8843(1)$ for $U(3)$ \cite{Unger:2011in} and $aT_c = 1.402(2)$ for $SU(3)$ \cite{forcrand:prl}, are determined assuming the mean field relation between the bare and renormalized anisotropy couplings, i.e. $aT_c = \gamma^2 a_t T_c$. Using our non-perturbative method for setting the anisotropy scale, the corresponding values for the $U(3)$ and $SU(3)$ critical temperatures, $a T_c = \xi(\gamma) a_t T_c$, deviate from those in the literature by $\approx 25\%$.}

The dependence of the energy density on the temperature, in $U(3)$ QCD, is given in Fig.~\ref{fig:energy} (right). The data points seem to fall on an universal curve, which deviates from SB (dotted line) at temperatures near $T_c$, and also at low temperatures. It is qualitatively consistent with the (analytical) mean field prediction in the large $N$ limit (solid line).\footnote{~In this mean field approach, the critical temperature is not easy to fix. For comparison with the $U(3)$ and $SU(3)$ data, we set the critical temperature to the $U(3)$ mean field value: $a T_c = 5/2$ \cite{aniso-mean-field}.}

The surprising deviation from SB at low temperatures may be due to finite size effects: the data points at the lowest temperatures require large $N_t$, but are computed for possibly not large enough values of $N_s$. Simulations with larger spatial volumes are required for a better control of the thermodynamical limit. On the other hand, the deviation from SB at high temperatures may be due to an UV cutoff effect, and simulations with larger values of $\xi$ are required in order to increase $N_t$, at fixed temperature. At intermediate values of the temperature, the energy density is the closest to SB, with a small discrepancy which may be due to a finite-$\xi$ effect: SB scaling is only expected to be exact in the $\xi\to\infty$ (continuous time) limit.

\begin{figure}[t]
\centering
\includegraphics[scale=.36]{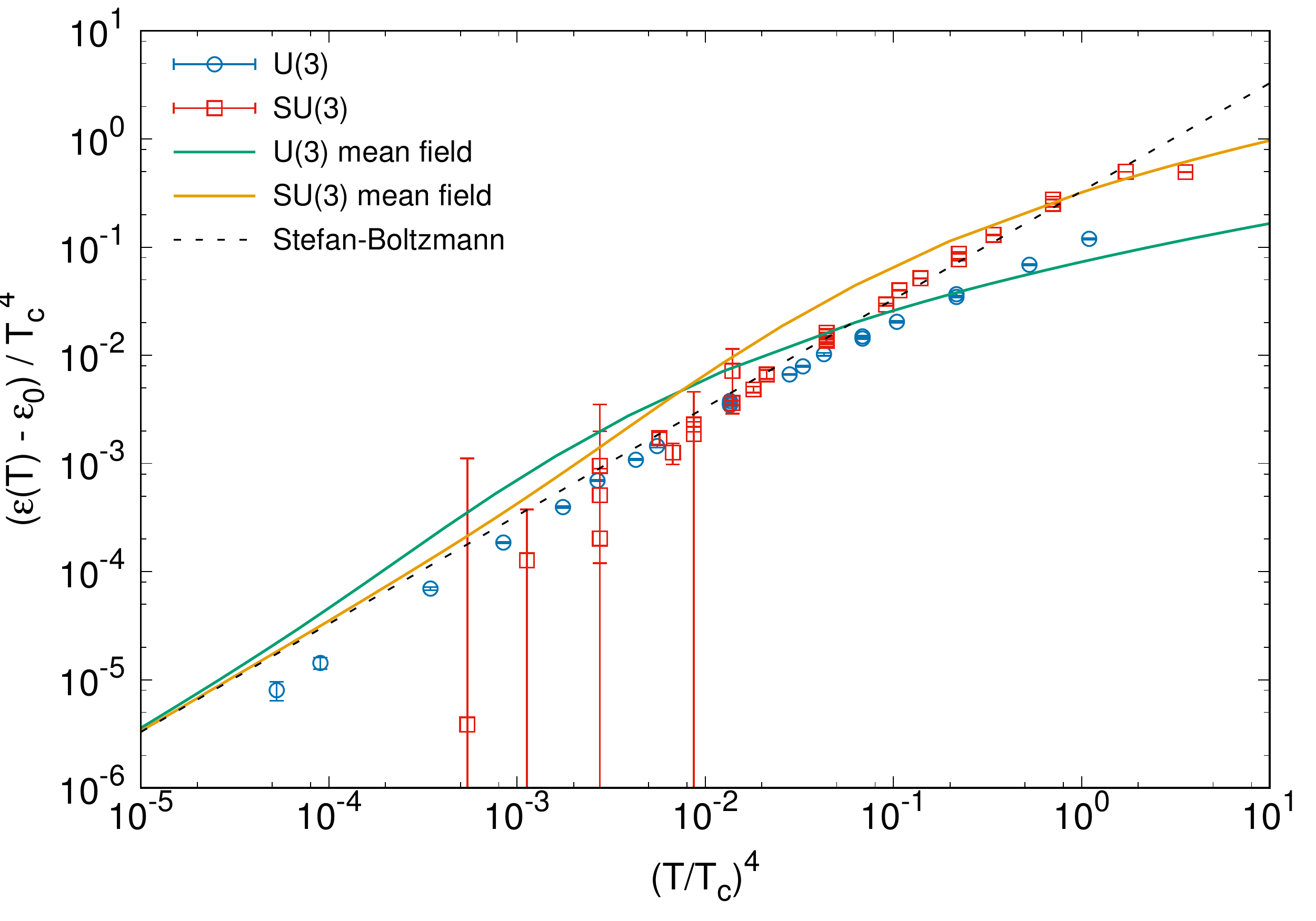}
\vskip -4mm
\caption{Comparison of the energy densities in $U(3)$ QCD (blue) and $SU(3)$ QCD (red), as a function of the temperature, in units of the critical temperatures of the respective chiral phase transitions. The large error bars in $SU(3)$ QCD, at low temperatures, are due to large fluctuations in the baryonic sign. The mean field curves are computed analytically in the large $N$ limit, and we set $a T_{c,U(3)} = 5/2$ and $a T_{c,SU(3)} = 5/3$ \cite{aniso-mean-field}.}
\label{fig:compare}
\end{figure}

Using the same approach, we have also computed the energy density as a function of the temperature in $SU(3)$ lattice QCD. A comparison between the $U(3)$ and $SU(3)$ theories is shown in Fig.~\ref{fig:compare}. The difference between the two cases is the additional contribution of baryon loops to the $SU(3)$ theory, which also introduces a sign problem, thus increasing the statistical error, especially for the large volumes required at low temperatures. 

At high temperatures, up to $T_c$, the $SU(3)$ energy density is consistent with SB, and consistently higher than that for $U(3)$, which we understand as being due to the thermal excitation of the extra baryonic modes, which contribute to the total pressure (and energy density).

In conclusion, the ideal pion gas is a reasonably good approximation for the low $T$ regime of $U(3)$ and $SU(3)$ lattice QCD. Deviations from it, associated with pion interactions and with the thermal excitation of massive hadrons, can be quantified and deserve further study. The study of the equation of state can also be extended to the cases of non-zero quark mass and of non-zero chemical potential.

%------------------------------------------------------------------------------------------------------------


\begin{thebibliography}{99}

\bibitem{Engels:1981ab}
  J.~Engels, F.~Karsch, H.~Satz,
  %``Finite Size Effects in Euclidean Lattice Thermodynamics for Noninteracting Bose and Fermi Systems,''
  Nucl.\ Phys.\ B {\bf 205} (1982) 239.
  %doi:10.1016/0550-3213(82)90387-X
  %%CITATION = doi:10.1016/0550-3213(82)90387-X;%%
  %121 citations counted in INSPIRE as of 26 Aug 2016
  
\bibitem{mdp}
  P.~Rossi, U.~Wolff,
  %``Lattice {QCD} With Fermions at Strong Coupling: A Dimer System,''
  {\em Nucl.\ Phys.} {\bf B248} (1984) 105; 
  F.~Karsch, K.~Mutter,
  %``Strong Coupling QCD At Finite Baryon Number Density,''
  {\em Nucl.\ Phys.} {\bf B313} (1989) 541.

\bibitem{worm}
  D.~H. Adams, S.~Chandrasekharan,
  %{\it {Chiral limit of strongly coupled lattice gauge theories}},
  {\em Nucl.Phys.} {\bf B662} (2003) 220. %--246. 
  [\href{http://arxiv.org/abs/hep-lat/0303003}{{\tt hep-lat/0303003}}].

%\cite{deForcrand:2009dh}
\bibitem{deForcrand:2009dh}
  P.~de Forcrand, M.~Fromm,
  %``Nuclear Physics from lattice QCD at strong coupling,''
  Phys.\ Rev.\ Lett.\  {\bf 104} (2010) 112005.
  %doi:10.1103/PhysRevLett.104.112005
  [\href{http://arxiv.org/abs/0907.1915}{{\tt arXiv:0907.1915}}].
  %[arXiv:0907.1915 [hep-lat]].
  %%CITATION = doi:10.1103/PhysRevLett.104.112005;%%
  %88 citations counted in INSPIRE as of 06 Sep 2016
  
%\cite{Unger:2011in}
\bibitem{Unger:2011in}
  W.~Unger, P.~de Forcrand,
  %``Continuous Time Monte Carlo for Lattice QCD in the Strong Coupling Limit,''
  PoS LATTICE {\bf 2011} (2011) 218.
  %[arXiv:1111.1434 [hep-lat]].
  [\href{http://arxiv.org/abs/1111.1434}{{\tt arXiv:1111.1434}}].
  %%CITATION = ARXIV:1111.1434;%%
  %8 citations counted in INSPIRE as of 28 Aug 2016
  
\bibitem{Chandrasekharan:2006tz}
  %\\
  S.~Chandrasekharan, F.-J. Jiang,
  %{\it {Phase-diagram of two-color lattice {QCD} in the chiral limit}},
  {\em Phys.Rev.} {\bf D74} (2006) 014506.
  %[\href{http://arxiv.org/abs/hep-lat/0602031}{{\tt hep-lat/0602031}}].

\bibitem{aniso-mean-field}
  N.~Bilic, F.~Karsch, K.~Redlich,
  %``Flavor Dependence of the Chiral Phase Transition in Strong Coupling QCD,''
  {\em Phys.\ Rev.}\ {\bf D45} (1992) 3228.
  %doi:10.1103/PhysRevD.45.3228
  %%CITATION = doi:10.1103/PhysRevD.45.3228;%%
  %44 citations counted in INSPIRE as of 26 May 2016

\bibitem{forcrand:prl}
  P.~de Forcrand, J.~Langelage, O.~Philipsen, W.~Unger,
  %``Lattice QCD Phase Diagram In and Away from the Strong Coupling Limit,''
  {\em Phys.\ Rev.\ Lett.\ } {\bf 113} (2014) 15,  152002.
  [\href{http://arxiv.org/abs/1406.4397}{{\tt arXiv:1406.4397}}].
  %[arXiv:1406.4397 [hep-lat]].

\end{thebibliography}
\end{document}